\documentstyle[aps,twocolumn,epsfig]{revtex}

\begin{document}

\title{Pressure drag in linear and nonlinear quantum fluids}

\author{T. Winiecki, J. F. McCann, and C. S. Adams }
\address{Dept. of Physics, University of Durham, Rochester Building,
South Road, Durham, DH1 3LE, England.}

\date{\today}
\draft
\maketitle

\begin{abstract}

We study the flow of a weakly-interacting Bose-Einstein condensate around
an obstacle by numerical solution of the Gross-Pitaevskii equation. 
We observe vortex emission and the formation of bow waves leading
to pressure drag. We compare the drag law with that of an ideal Bose gas, 
and show that interactions reduce the drag force. This reduction can be 
explained in terms of a `collisional screening' of the obstacle.

\end{abstract}


A central issue in fluid flow concerns the origin of resistance or drag. 
In a viscous fluid, shear stresses induced by friction at a surface
lead to {\it skin drag}. In an ideal fluid or superfluid, the effects of shear
stress vanish, but normal stresses induced by pressure gradients
across an obstacle still produce {\it pressure drag}.\\
 
The recent experimental breakthroughs allowing the production dilute Bose-Einstein 
condensates \cite{ande95} has stimulated new interest in the flow 
and modes of excitation of quantum fluids \cite{dalf98}. Dilute 
Bose-Einstein condensates are compressible nonlinear quantum fluids, 
with the attractive feature that the time-evolution can be accurately 
described by a nonlinear Schr\"odinger equation (NLSE), known as the 
Gross-Pitaevskii equation, allowing direct quantitative 
comparison between theory and experiment \cite{dalf98}.
One could envisage an experiment to measure the pressure drag in a dilute 
Bose-Einstein condensate by studying the flow past an obstacle such as a 
far-detuned laser beam \cite{jack98} or a foreign condensate species 
\cite{hall98,jack99}. From numerical solution of the NLSE, it is known that 
in a uniform flow there is a critical velocity $v_{\rm c}$ for the onset of 
drag \cite{fris92}, however, the drag coefficient and exact dependence on the 
flow velocity remain unknown.\\

In this paper, we address these issues by simulating the flow 
of a weakly-interacting Bose-Einstein condensate around penetrable and 
impenetrable objects. We show that for a flow velocity $v$ larger
than the critical velocity ($v>v_{\rm c}$), the 
drag force varies quadratically with $v$, similar to an 
ideal Bose fluid. The principle effect of interactions is to reduce 
drag, which we explain in terms of a `collisional screening' of 
the object. Also, we show that the effective size of the obstacle is increased 
due to the healing length of the fluid, and  consequently the drag is non-zero 
even for point-like objects.\\ 

The NLSE in a uniform potential may be written as, 
\begin{equation}
{\rm i} \hbar \partial_t \psi = - \frac{\hbar^2}{2m} \nabla^2 \psi 
+ V\psi + C \vert \psi \vert^2 \psi~,
\label{eq:NLSE}
\end{equation}
where the nonlinear coefficient, $C$, is determined by the strength
of particle interactions, and $V$ is the object potential.
This equation is equivalent to two hydrodynamic equations corresponding to 
the conservation of mass and momentum, respectively,
\begin{equation} 
\partial_t \rho + \partial_i J_i = 0,~{\rm and}~~~
\partial_t J_i + \partial_j T_{ij} = 0~,
\end{equation}
where 
\begin{equation} 
\rho = m\psi^* \psi,~{\rm and}~~~
J_i = \frac{\hbar}{2{\rm i}}(\psi^* \partial_i \psi - 
\psi \partial_i \psi^*)~,
\end{equation}
are the mass density and momentum flux, and 
\begin{equation} 
T_{ij} = {1 \over 2} \delta_{ij}C\vert\psi\vert^4 
+\frac{\hbar^2}{4m}(\partial_i \psi^* \partial_j \psi
-\psi^* \partial_{ij}\psi 
+{\rm c.c.}),
\label{eq:stress}
\end{equation} 
is the stress tensor. The instantaneous drag force on an obstacle is given by
\begin{equation}
F_k(t)=\int_S {\rm d}s_j~T_{jk}(t)~, 
\label{eq:force1}
\end{equation}
where $S$ defines the surface of the obstacle and ${\rm d}s_j$ is
an element of $S$ in the direction of the outward normal. For a linear fluid
($C=0$) and an impenetrable cylindrical obstacle of radius $R$, the drag
law may be derived analytically: for high velocity or large object size,
($v\gg\hbar/mR$) the force approaches the classical limit, 
\begin{equation}
F=\frac{8}{3}\rho_0 Rv^2~,
\label{eq:f_particle}
\end{equation}
where $\rho_0$ is the background mass density; 
whereas, for low velocity or small objects ($v\ll\hbar/mR$),
\begin{equation}
F=~\alpha(mvR/\hbar)\rho_0 \hbar v~,
\label{eq:f_wave}
\end{equation} 
where  
$\alpha(\zeta)=16/\pi^2\zeta^2\vert H_0^{(1)}(\zeta)H_1^{(1)}(\zeta)\vert^2$
is only weakly-dependent on velocity: $H_\nu^{(1)}$ is the Hankel function.
In this case, often encountered in acoustic scattering,
the wave properties dominate and the effective object
size is proportional to the wavelength. Inserting $R\sim\lambda=h/mv$ into
Eq.~(\ref{eq:f_particle}), implies
$F\sim \rho_0 hv$, consistent with Eq.~(\ref{eq:f_wave}).\\

For a weakly-interacting condensate ($C>0$), the drag force must be 
evaluated numerically. For the numerical solution, we adopt the reduced units, 
$\tilde{t}= t/(\hbar/n_0C)$, $\tilde{y} = y/(\hbar/\sqrt{mn_0C})$,
and $\tilde{\psi}=\psi/\sqrt{n_0}$, where $n_0$ is the in-flow number density. 
In reduced units, distance $\tilde{y}$, and velocity $\tilde{v}$ are 
measured in terms of the healing length, $\xi=\hbar/\sqrt{mn_0C}$, and
the speed of sound, $c=\sqrt{n_0C/m}$, respectively; and Eq.~(\ref{eq:NLSE})
becomes
\begin{equation} 
{\rm i}\partial_{\tilde{t}}\tilde{\psi}=-{1\over 2}\nabla^2\tilde{\psi}+
\tilde{V}\tilde{\psi} + \vert\tilde{\psi}\vert^2~.
\end{equation} 
This equation is solved in 2D for both penetrable and impenetrable 
objects \cite{numerics}. A uniform flow in the $-y$-direction 
is imposed by multiplying the stationary solution by a phase
gradient, ${\rm e}^{-{\rm i}\tilde{v}\tilde{y}}$. In principle, the 
instantaneous drag force can be determined by numerical integration of the 
stress tensor at the surface of the obstacle, Eq.~(\ref{eq:force1}). However, 
for impenetrable objects, the finite grid size introduces errors in the 
differencing approximations to the surface derivatives of $\psi$. And for 
penetrable objects, this procedure is complicated 
because the fluid-object boundary is ill-defined. The numerical integration 
can be greatly simplified by recognizing that the time-averaged drag force 
must be equal to the back-action on the fluid, i.e., the instantaneous force,
\begin{equation}
F_k(t)=-\int_\Gamma {\rm d}s_j~T_{jk}(t) - 
\frac{\partial}{\partial t}\left[\int_A {\rm d}A~J_k(t)\right]~, 
\label{eq:force2}
\end{equation}
where $\Gamma$ defines the outer border of a simply-connected region of fluid, $A$, 
encircling the object. The second term, which corresponds to
the rate of change of the fluid momentum 
within $A$, averages to zero, if the flow velocity remains constant.\\

\begin{figure}
\epsfig{file=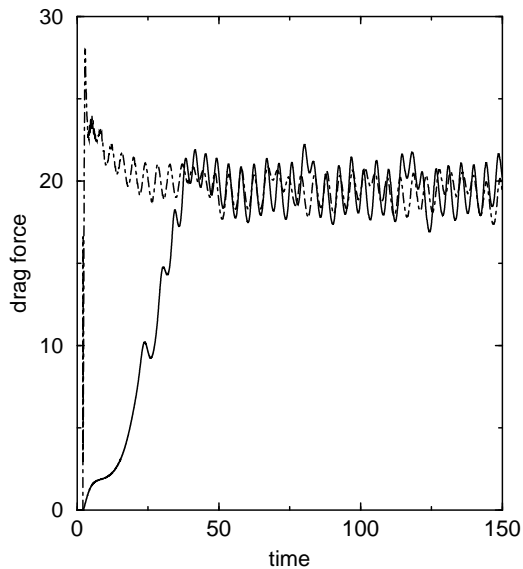,clip=,width=7cm}
\caption{The instantaneous force, $F_y$ (in units of
$\hbar\sqrt{n_0^3C/m}$), on an obstacle placed in a
nonlinear quantum flow with velocity $\tilde{v}=1.5$, 
as a function of the reduced time, $\tilde{t}$. 
The two curves corresponds to switching on the flow gradually (solid) or
instantaneously (dashed). The oscillations are produced by the 
periodic emission of vortex pairs. Although the instantaneous drag depends
on the initial conditions, the time-averaged drag does not.}
\label{fig:1}
\end{figure}

Eq.~(\ref{eq:force2}) may be used to calculate the drag for both 
penetrable and impenetrable objects. Fig.~(\ref{fig:1}) shows a plot of the 
instantaneous drag, $F_y(\tilde{t})$, on an impenetrable cylinder with 
radius $\tilde{R}=5$, in a flow with velocity $\tilde{v}=1.5$.
Initially, the force is dominated by transients which depend
on how the flow is turned on. For an instantaneous turn-on, 
reflections from the obstacle produce
sound waves, which are subsequently absorbed at the edges of the 
box \cite{numerics}. However, for longer times the time-averaged drag is 
independent of the initial conditions. Also, the time-averaged force
is independent of the integration path $\Gamma$, and for barrier
height, $\tilde{V}>1$, only weakly dependent on the penetrability 
of the obstacle.
 
\begin{figure}
\epsfig{file=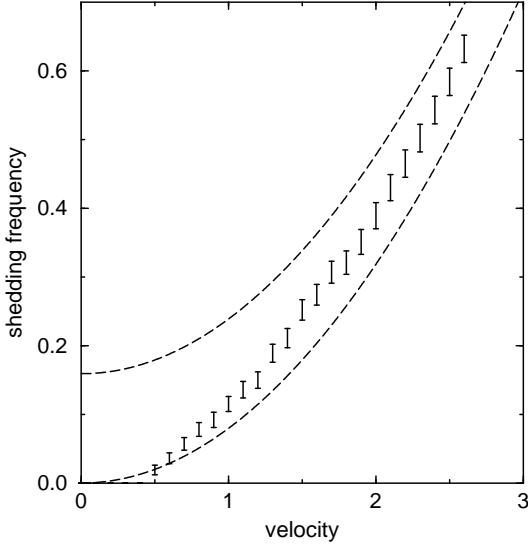,clip=,width=7cm}
\caption{The vortex shedding frequency, $f$ (in units of $n_0C/\hbar$),
as a function of the flow velocity, $\tilde{v}$. The numerical results 
lie between the dashed lines, $\tilde{v}^2/4\pi$ and $(\tilde{v}^2+2)/4\pi$, 
which correspond to the upper and lower bounds predicted by considering the 
phase-slip between the main flow and an almost stationary wake.
The error bars reflect the fluctuations in the vortex shedding frequency.}
\label{fig:2}
\end{figure}

The oscillatory behaviour of the instantaneous drag is produced by the 
periodic emission of vortex pairs. The vortex shedding frequency 
follows from the phase-slip between the main flow
and the almost stationary wake behind the obstacle: the wavefunctions 
for the flow and the wake may be written as 
$\tilde{\psi}={\rm e}^{-{\rm i}(1+\tilde{v}^2/2)\tilde{t}}
{\rm e}^{-{\rm i}\tilde{v}\tilde{y}}$ 
and $\tilde{\psi}=\tilde{n}^{1/2}{\rm e}^{-{\rm i}\tilde{n} \tilde{t}}$, 
respectively, where $\tilde{n}$ is the mean density behind 
the obstacle ($0<\tilde{n}<1$, decreasing at higher velocity). 
A vortex pair is emitted each time 
the phase difference accumulates to $2\pi$, giving a shedding frequency, 
$f=(1+\tilde{v}^2/2-\tilde{n})/2\pi$. Fig.~\ref{fig:2} shows a comparison
between the numerical results and the phase-slip model. The shedding frequency 
lies between the upper and lower limits set by the density, except at 
low velocity, where the shedding frequency falls to zero.

\begin{figure}
\epsfig{file=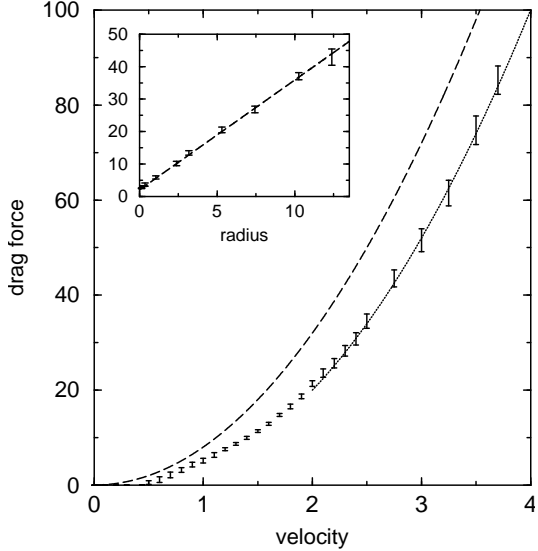,clip=,width=7cm}
\caption{The time-averaged drag force, $\overline{F}_y$ (in units of
$\hbar\sqrt{n_0^3C/m}$) as a function of flow velocity, $\tilde{v}$, 
for a impenetrable cylinder with radius $\tilde{R}=3$. The error bars indicate the 
magnitude of residual fluctuations in the time-averaged drag.
The drag law for a linear fluid is indicated by the dashed line. 
The effect of the non-linearity is to reduce
the drag which can be understood in terms of a collisional screening
of the object. The reduced drag predicted for a screened object is
indicated by the dotted line. The force as a 
function of the object size for $\tilde{v}=1.5$ is shown inset. Note that, 
the drag is non-zero even for objects much smaller than the healing length.
}
\label{fig:3}
\end{figure}

The subsequent motion of the vortices is complex: the first pair is overtaken by
subsequent pairs and becomes `trapped' behind the obstacle.
However, as apparent in Fig.~1, the vortices do not 
contribute directly to the time-averaged force, their primary
role is to allow the separation of the wake, which then results in
a pressure gradient across the obstacle.  
Fig.~\ref{fig:3} shows the time-averaged drag as a function of flow
velocity. The error bars correspond to the residual fluctuations
after averaging. The drag curve for the linear fluid, 
Eqs.~(\ref{eq:f_particle}) and (\ref{eq:f_wave}), is shown as a dashed line.
One sees that the main effect of particle interactions is to 
reduce the drag.  Below a critical velocity, $v_{\rm c}$,
the drag in the nonlinear fluid falls to zero. In this region, 
the solutions of the NLSE are time independent and symmetric. The numeric value 
of $v_{\rm c}$ depends on the object shape and penetrability.
For an impenetrable cylinder, we find $v_{\rm c}=(0.45\pm 0.01)c$, 
consistent with previous work \cite{fris92,huep97}. 
The critical velocity is larger for an impenetrable square barrier, 
and lower for a penetrable Gaussian object.

Above the critical velocity, a pressure drag appears with a velocity 
dependence similar to the linear fluid. In fact, the force can be predicted 
accurately by a semi-classical modification of the linear drag law, i.e.,
\begin{equation}
F=\frac{8}{3}\rho_0 R'v'^2~,
\label{eq:f_non-linear}
\end{equation}
where $R'$ and $v'$ are an effective object radius and flow velocity,
respectively.  A plot of the drag force as a function of
object size, Fig.~\ref{fig:3}(inset), indicates that the effective object 
size is extended by the healing length $\xi$ of the fluid.
For $\tilde{v}=1.5$, we find that $R'=R+0.6\xi$.
An important consequence of this result is that
the force does not vanish for small objects, $R\ll\xi$, 
e.g. a point defect in a superconductor.\\

\begin{figure}
\epsfig{file=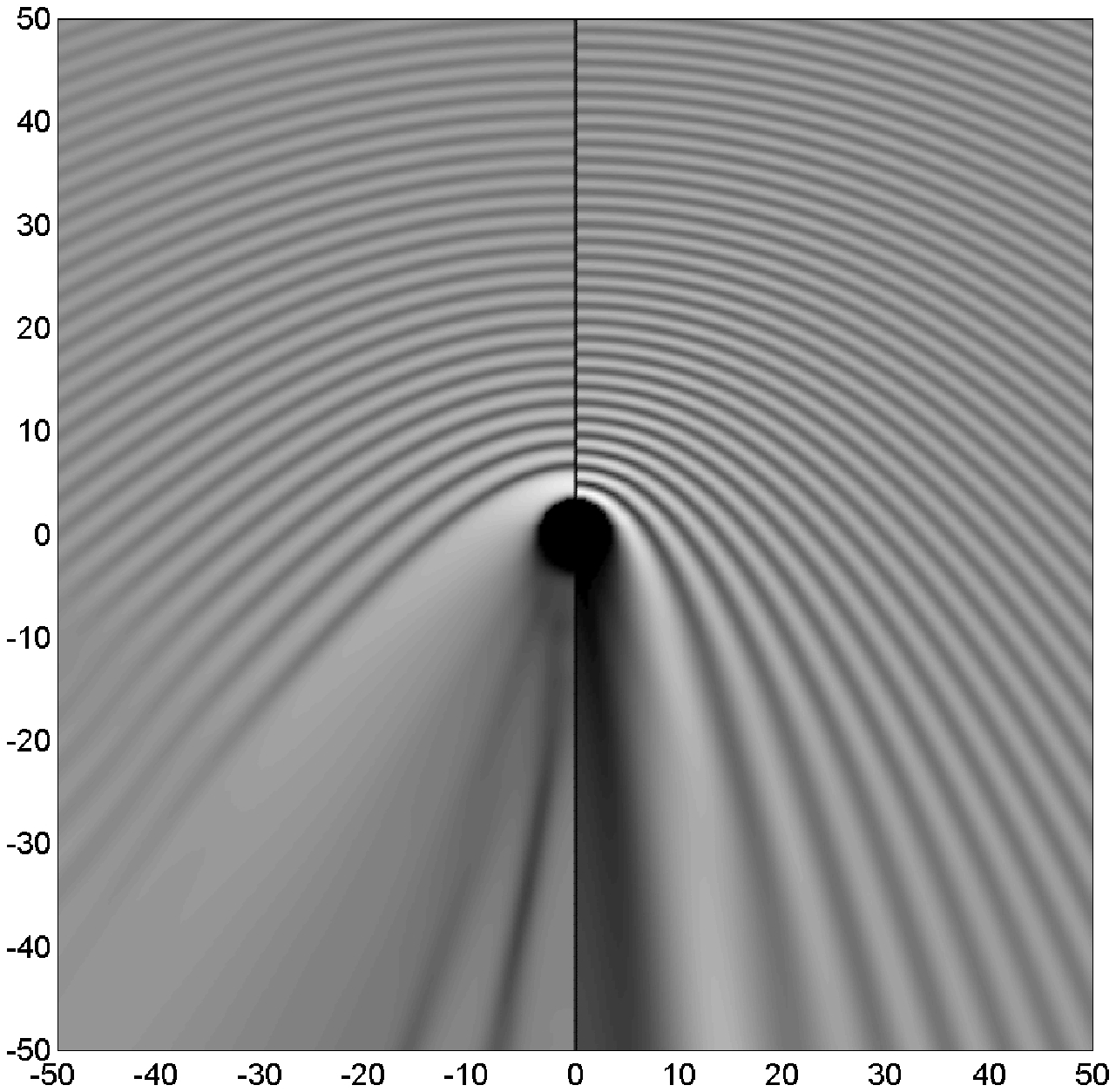,clip=,width=7.5cm,bbllx=110,bblly=230,bburx=485,bbury=610}
\caption{The time-averaged density distribution for $\tilde{v}=2.0$ with (left) and 
without (right) interactions. The obstacle is an impenetrable cylinder 
with radius $\tilde{R}=3$ centred at the origin, and the flow
is from top to bottom. The repulsive particle 
interactions tend to smooth density variations,
thereby reducing the quantum pressure experienced by the obstacle. 
For the nonlinear fluid (left), the dark line in the wake, close to the axis 
of symmetry, corresponds to the vortex street.}
\label{fig:4}
\end{figure}

The effective flow velocity, $v'$, may be estimated by considering how the
flow is modified by the collisional mean-field. Above the critical velocity, 
incoming waves are reflected by the obstacle producing a standing wave or bow wave
(see Fig.~\ref{fig:4}), with a maximum density variation (relative to
the background flow) of between $-n_0$ and $+3n_0$. In a semi-classical 
treatment, an oscillatory
potential slows incoming particles by an amount corresponding 
to half the maximum barrier height, i.e., the effective flow 
velocity, $v'^2=v^2-\Delta c^2$, where $0<\Delta<3$. 
The dotted line in Fig.~3 is a plot of Eq.~(\ref{eq:f_non-linear}) using 
values of $\Delta$ obtained from the numerical solution. 
This `collisional screening' model is accurate at high velocity, 
but less so at lower velocity, where the semi-classical particle 
treatment begins to break down.\\
  
Fig.~4 show a comparison of the time-averaged density
distribution for nonlinear (left) and linear (right) quantum fluids.
One sees that the finite compressibility of the nonlinear fluid tends to suppress
large density fluctuations leading to a smoothing of the 
standing wave in front of the obstacle. 
Also far downstream, the direction of the bow waves approaches the 
Mach angle, $\alpha=\sin^{-1}(c/v)$.
For the linear fluid, $c=0$ and $\alpha=0$, i.e., the bow
waves runs adjacent to the geometric shadow behind the obstacle. 
In the nonlinear fluid, the `shadow' is far less pronounced: the dark streak
in the wake, close to the axis of symmetry, corresponds to the vortex street. 
 
To summarize, we have solved the NLSE equation in 2D to
simulate the flow of a weakly-interacting Bose-Einstein
condensate around an obstacle. We observe vortex emission
and the formation of bow waves leading to a pressure drag. 
We find that the drag force is proportional to the screened
energy of the flow, and to the object cross-section
extended by the effect of fluid healing. 

\acknowledgements
TW is supported by the Studienstiftung des Deutschen Volkes.\\


\begin{references}
\bibitem{numerics}
The time evolution is evaluated using a semi-implicit Crank-Nicholson formula,
and applying the split-operator alternating-direction method for spatial 
differencing. The resulting nonlinear equations are solved using Newton's method. 
The calculations were performed using a time step, $\tau=0.01$, and a box 
size of 200$\times$200 healing lengths, divided into 800$\times$800 points. 
The box consists of a central region surrounded by an absorbing border of
width 80 points. At the edges, boundary conditions which maintain a uniform flow
are applied.
\end{references}
\end{document}